\documentclass[letterpaper,12pt]{article}
\usepackage[
   top=25mm,
   bottom=25mm,
   left=25mm,
   right=25mm]{geometry}
\usepackage{color}
\usepackage{enumitem}
\usepackage{amsmath}
\usepackage{graphicx}
\graphicspath{ {./} }

\providecommand{\keywords}[1]
{
  \tiny	
  \textbf{\textit{Keywords---}} #1
} 

\begin{document}

\title{Combining Blotto Networks and Voter Models to Simulate Voter Behavior in Response to Competitive Election Spending}
\author{Renee Jerome, DePaul University}
\date{\today}

\maketitle

\small
Blotto Games are regularly used to simulate spending in political campaigns because they represent the behavior of two competing players distributing finite resources over a finite number of battlefields or districts. By representing these districts as Voter Networks, we hope to simulate the behavior of voters in response to competitive campaign spending and identify Blotto strategies that consider both network data as well as a straightforward Blotto approach. Winning a network in the Blotto game is not enough to guarantee that the voters will sway towards a specific candidate, particularly in highly polarized graphs. It may be possible to identify winnable battlefields and assign these more value; alternatively, knowing the strategy of an opponent, it may be possible to prevent them from obtaining the battlefields they consider valuable. 
\newline
\normalsize
\section{Introduction}
\quad Advertising, or propaganda, has long played a powerful role in shaping public opinion on both political and nonpolitical matters. In an increasingly rapid media cycle with near-instant access to voters or consumers, this role can only become more important. In many cases, advertisers or other organizations now also have access to data about how a population is interconnected, such as social media connections or shared contexts such as home, school, workplace, etc. This depth of information may be able to inform how propaganda or advertising is disseminated as part of a media campaign and which populations are of most value to the advertiser, thus allowing for more targeted dissemination of resources to greater effect.  
\newline \newline
\quad In the past, the Voter Model has been explicitly used to model the impact of propaganda on a dynamic, interconnected population, and certain factors have been identified that influence the behavior of voters when under outside influence \cite{Manohara}. The Blotto Game has also been explicitly used to study information wars between two opposing parties, whether in regards to a political issue or advertising war \cite{Watts}. 
\newline \newline
\quad There are many variations of the Voter Model that are used to study different components of voter behavior. One component is resistance to change; some voters, usually termed Zealots, will not change color under any circumstances \cite{Mobilia}. Even just a small number of Zealots can prevent convergence, or even prevent a strong majority from forming. Similarly, Heterogenous Models set each Voter’s rate of change individually, representing different degrees of changeability for different individuals \cite{Masuda}. A Voter’s position can also be modeled on a sliding scale rather than as a binary state \cite{Congleton}. In this conception, a Median Voter can be identified for whom half of the nodes have a position that is above them on the scale and half have a position that is below them. The behavior of the Median Voter both influences and is influenced by the state of the larger graph. 
 \newline \newline
\quad Both the graph theory behind the Voter Model and the game theory aspects of the Blotto Game are relevant to the behavior of voters or consumers when they are under the influence of competing propaganda campaigns, and for this reason both are useful to understand the most effective spending strategy. In this project, we seek to combine the two problems into a Voter-Blotto Game and examine what components of the graph most effect its value in the eyes of the competing players.  
\section{Defining the Component Algorithms}
\subsection{Defining the Voter Model}
\quad For the purposes of this paper, we define the Voter Model as in \cite{Manohara}: an unweighted, undirected network of connected nodes. Each node represents a voter, and each edge represents a path of influence. The nodes are colored either red or blue to represent the node’s current loyalty. Below is an example of a small Voter Model representing five voters, three of which are blue and two of which are red. 
\newline
\includegraphics[scale=1]{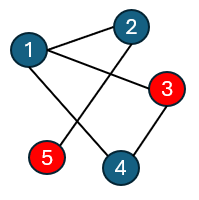}\newline
\quad We refer to polarization as the degree to which a graph skews towards the more populous color. In this case, at 60\% blue and 40\% red, we would say that this graph has a polarization of 60\%. In each round of the Voter Model, we do the following steps.
\newline \newline
\begin{enumerate}
\item
Iterate over each node.
\begin{enumerate}[label=(\alph*)]
\item
Count how many red neighbors and how many blue neighbors the node has.
\item
If it has more red neighbors, change its color to red.
\item
If it has more blue neighbors, change its color to blue.
\item 
If it has an equal number of red and blue neighbors, do not change the color.
\end{enumerate}
\end{enumerate}

\quad This is repeated until the graph no longer changes color. This is referred to as converging. A graph may converge to entirely red or entirely blue (100\% polarization), or it may converge to a state that contains both red and blue, or it may not converge at all and will cycle indefinitely through some number of states. We know the polarization and size of a graph influence how long it will take to converge, with smaller graphs of higher polarization converging more quickly. However, the shape of a graph also heavily influences how long it will take to converge, or if it will converge at all. 
\newline \newline
\quad For that reason, we must find a shape that will accurately represent a pool of voters and how people group together and form connections in real life. To do this, we are using a Small World Graph, defined by Watts and Strogatz in 1998 \cite{Watts}. Many real-world phenomenon can be described with Small World Graphs, including social media connections, interactions between budding yeast cells, and, most importantly, how groups of people form clusters and connections. A Small World Graph is defined as having a short average path length, meaning that for any two randomly chosen nodes in the graph, the length of the shortest path between them will be, relatively speaking, on average fairly short. For example, in a graph of 12 nodes, the average path length might be approximately 2. It also has a high clustering coefficient. The graph is created by connecting each node to its k nearest neighbors, and then iterating over each edge and, with probabiliy p, potentially transferring the edge to a random node. This means that a graph with a higher k will be denser, and a graph with a higher p will be more random. 
\newline \newline
\quad As defined in “A Generalisation of Voter Model: Influential Nodes and Convergence Properties” \cite{Manohara}, we are also applying thresholds, where a node won’t flip until a certain percentage of its neighbors are one color; for example, it won’t flip until at least 70\% of its neighbors are of the opposing color. We start at a threshold of 50\% and multilpy by 1.05 every round; this simulates a time-based flip rate, approximating the tendency of voters to be more easily influenced at the beginning of an election cycle and less easily influenced closer to voting day.

\subsection{Defining the Blotto Game}
\quad A Blotto Game is a game theory problem that can be used to study distribution of a finite amount of resources over multiple battlefields by opposing players. “Blotto game in a propaganda battle”\cite{Podlipskaia} explicitly models an information war as a Blotto Game. We define the game similarly. There are two players and n battlefields, and each player distributes a certain amount of their resources to each battlefield without being able to see the other player’s distribution. Whichever player has distributed more resources to a battlefield wins that field; whichever player wins more battlefields wins the round. This is analogous to opposing campaigns distributing advertising or other resources across multiple voting districts, trying to outweight their competitor’s advertising in as many districts as possible while working with a finite budget and without knowing the exact distribution of their opponent’s advertising budget. 
\newline \newline
\quad While both the Voter Model and the Blotto Game have been heavily studied in reference to information warfare and the impact of propaganda on a population, we seek here to combine the two to understand how the game theory aspect and the behavior of the graph interact. 

\subsection{Defining the Voter-Blotto Game}
\quad To combine these two ideas, we set up a Blotto Game where the battlefields are Voter Models. Each network has different shape and connectivity. The players are now referred to as the Red Player and the Blue Player; their objective is to have as many nodes in their respective color as possible. Each player distributes their resources across the battlefields (networks), and the player that wins a battlefield is given a “boost” when calculating the next color for each node. As before, we iterate over each node and count how many red and blue neighbors it has; however, before comparing the values, we multiply the winning color by some constant larger than 1. The constant is selected via a logistic function based on the ratio of the winning player’s resources to the losing player. 

\[C = \frac{1}{1 + e^{-k(r_1/r_2 - 1.5)}}\]

Where C is the constant boost, k controls the sharpness of the curve, and the r values refer to the amount of resources distributed by the two players. 

\newpage
Our process, then, looks something like this: \newline

\includegraphics[width=\textwidth]{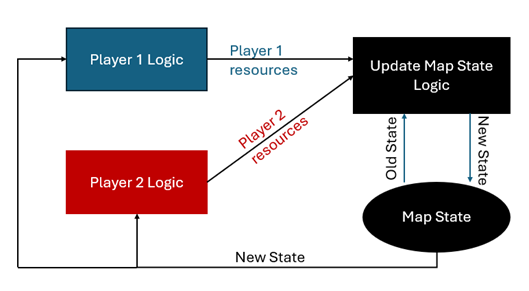}\newline

\quad The players, using different logic strategies, will distribute a certain amount of resources to each battlefield in the map. The map update logic – the Voter Model with the boost from the Blotto Game – will update the map state. This new state is then passed to the player logic models so they can determine what resources to play in the next round. Every time the map is updated, the players will again assess the map and determine how many resources they want to distribute. In a game of 100 rounds, each player will make 100 "decisions" about where to distribute their resources. 

\section{Observations}

\quad We examined various features of the voter model in order to understand what factors infludnece the winnability ofa  graph. 2000 small world graphs with 300 nodes each were randomly generated, and then allowed to cycle to convergence. The value of k was randomized between 5 and 10; the polarization cycled from 45\% blue to 55\% blue in increments of 1\%, as values outside this range had very consistent convergence to red and blue respectively; and p was a random value between 0 and 1. Because the data is randomly generated, it is not gauranteed to be continuous for any given feature. We also gathered other data about the graphs, including the density, the diameter, the average path length, the average clustering, the maxmial degree of any node, the number of nodes of each color with this degree, the number of nodes in the top ten most connected nodes that are red or blue, the total degree of these nodes, and the total red degree and total blue degree across the entire graph.  
\newline \newline
\quad We analyzed a selection of these features using heatmaps. For each comparison, the first graph represents the convergability of the graph, i.e. the number of rounds to convergence. The simulation cut off after 100 rounds. The data was sorted into “buckets”, approximately 40 for each feature, and the number of rounds to convergence was then averaged within the bucket to get an approximate value for this region of the heatmap. Black represents a hole in the data, and columns that had no valid data were removed from the map. 
\newline \newline
\quad Some comparisons also include a second, bicolor graph that is both red and blue. For this grpah, a new value was created called “average color.” Graphs that converged to blue were counted as -1 and graphs that converged to red were counted as +1. Thus, when the data points are summed and averaged, regions of the heatmap that tend to go to blue will have a negative sum and regions of the heatmap that tend to go red will have a positive sum. For graphs that are equally or near-equally split, the tile will display white or pale blue/red. Graphs that did not converge do not affect the average color. 

\subsection{Average Clustering vs Polarization}
\includegraphics[width=\textwidth]{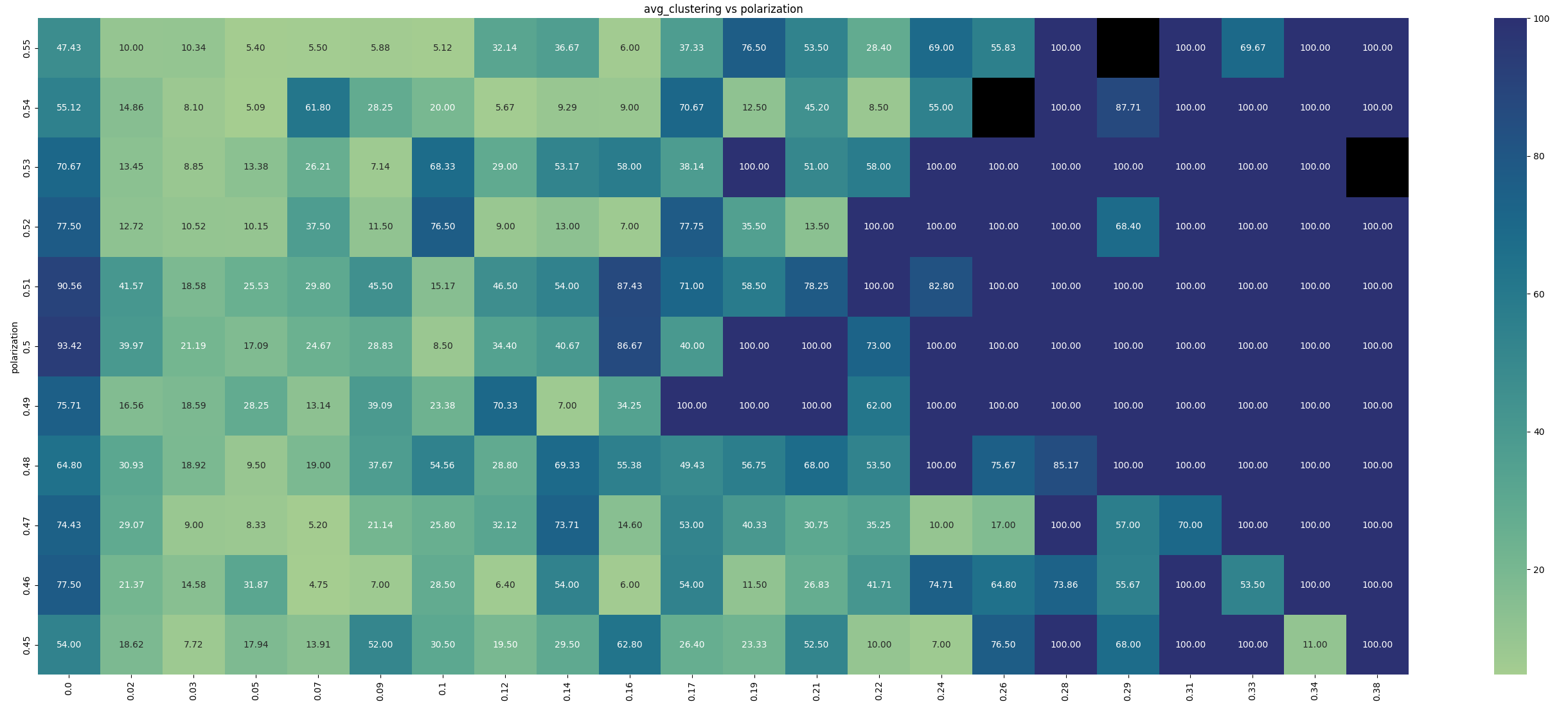}\newline
\includegraphics[width=\textwidth]{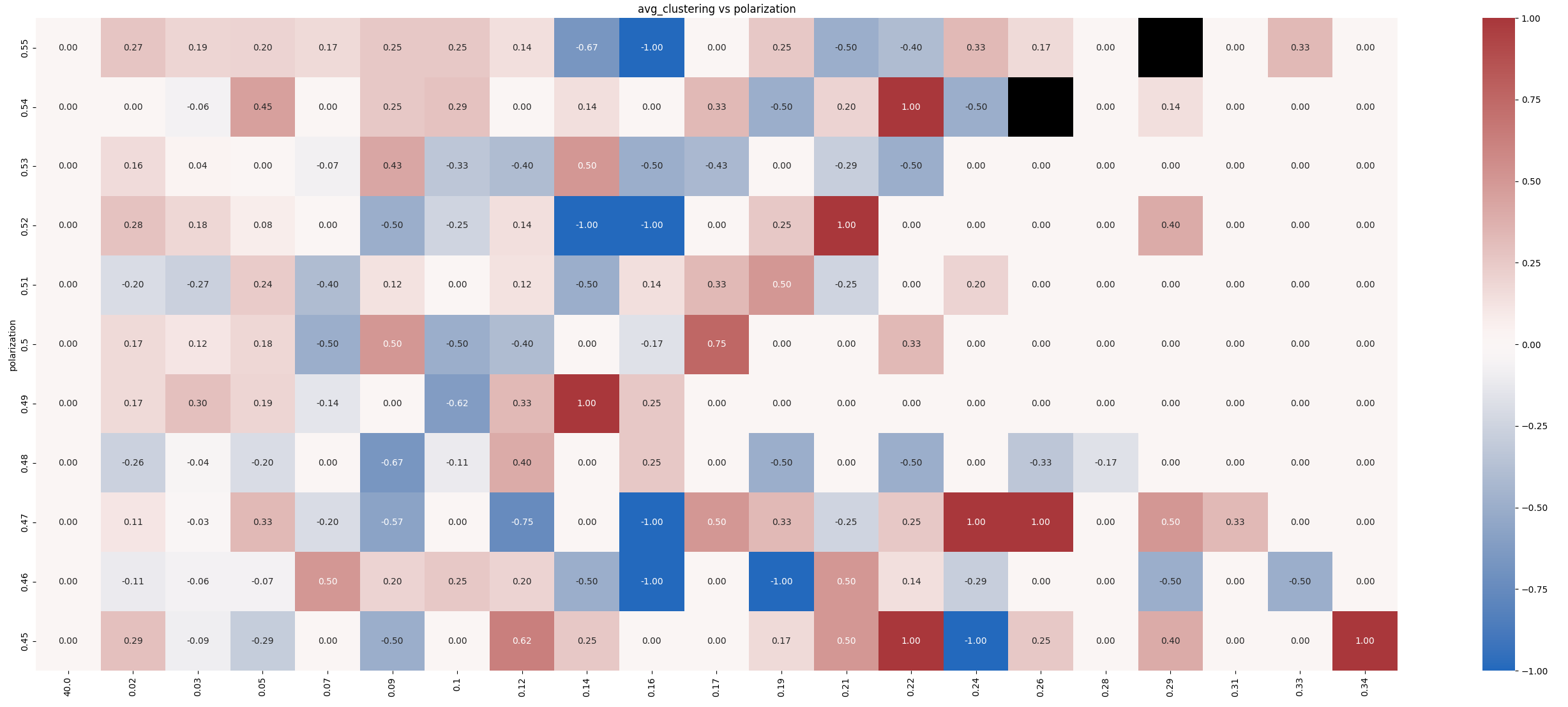}\newline
From the first map, we can see that as the clustering increases, the likelihood of requiring 100 or more rounds to converge increases as well. At the lefthand side of the graph, where the clustering is low, the majority of the graphs are able to converge in a reasonable number of rounds. But as we move to the right, more and more graphs require 100+ rounds, starting with those at the center (polarization closest to 50\%) and increasing. Thus, we can extrapolate that as a graph becomes more clustered, it is more difficult to force the graph to converge. 
\newline \newline
In the second, bicolor map, we see a similar phenomenon occurring with the white tiles. This graph represents the average color of convergence, and a white tile indicates an average of 0, meaning it is equally likely to be red or blue; or, that the graph failed to converge to either color. It does not appear that the clustering strongly effected whether the graph would be red or blue; rather, it led to many cases of failing to converge

\subsection{Density vs Polarization}
\includegraphics[width=\textwidth]{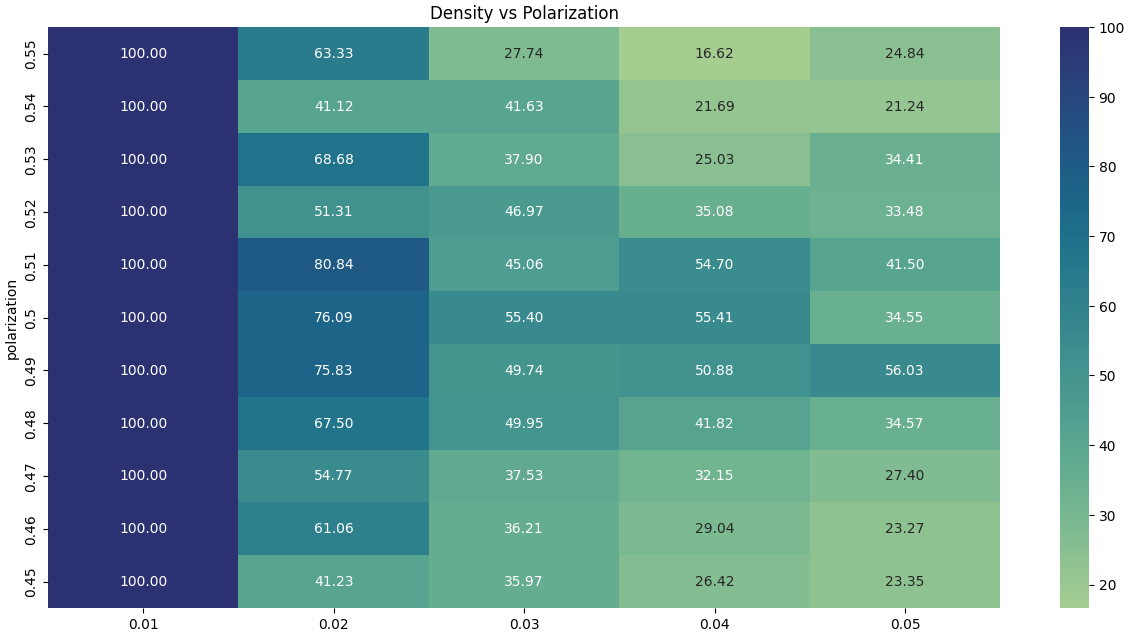}\newline
\quad We see a clear trend towards faster convergence moving towards a higher density. This does not seem to be strongly correlated to the polarization; the graphs in the center may take slightly longer to converge, particularly in the middle region of the graph, but the more distinct trend is from left to right. Increased density increases the rate of convergence at any polarization. Again, the second map does not indicate any noticeable correlations.

\subsection{Number of Neighbors k vs Polarization}
\includegraphics[width=\textwidth]{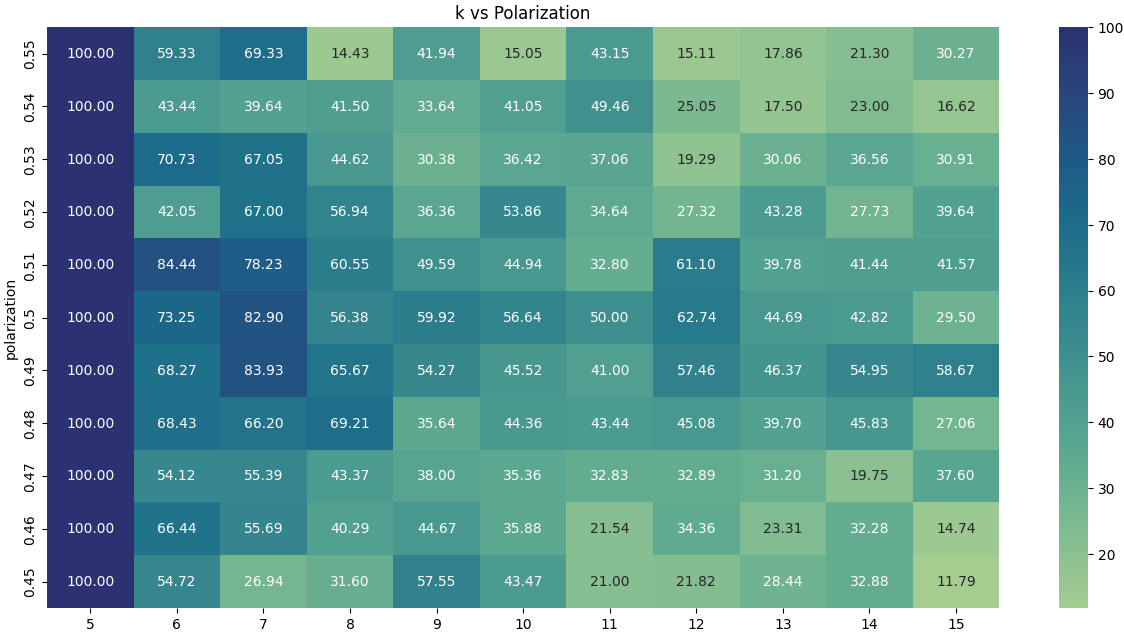}\newline
\quad Recall that k represents the number of neighbors that a node in a small world graph is connected to before applying the randomization element. We again see a trend from left to right, but it is less distinct than in either previous graph.  Because an increased k is likely to correlate to an increased density, this follows from our observations of Density vs Polarization. 

\subsection{Probability p vs Polarization}
\includegraphics[width=\textwidth]{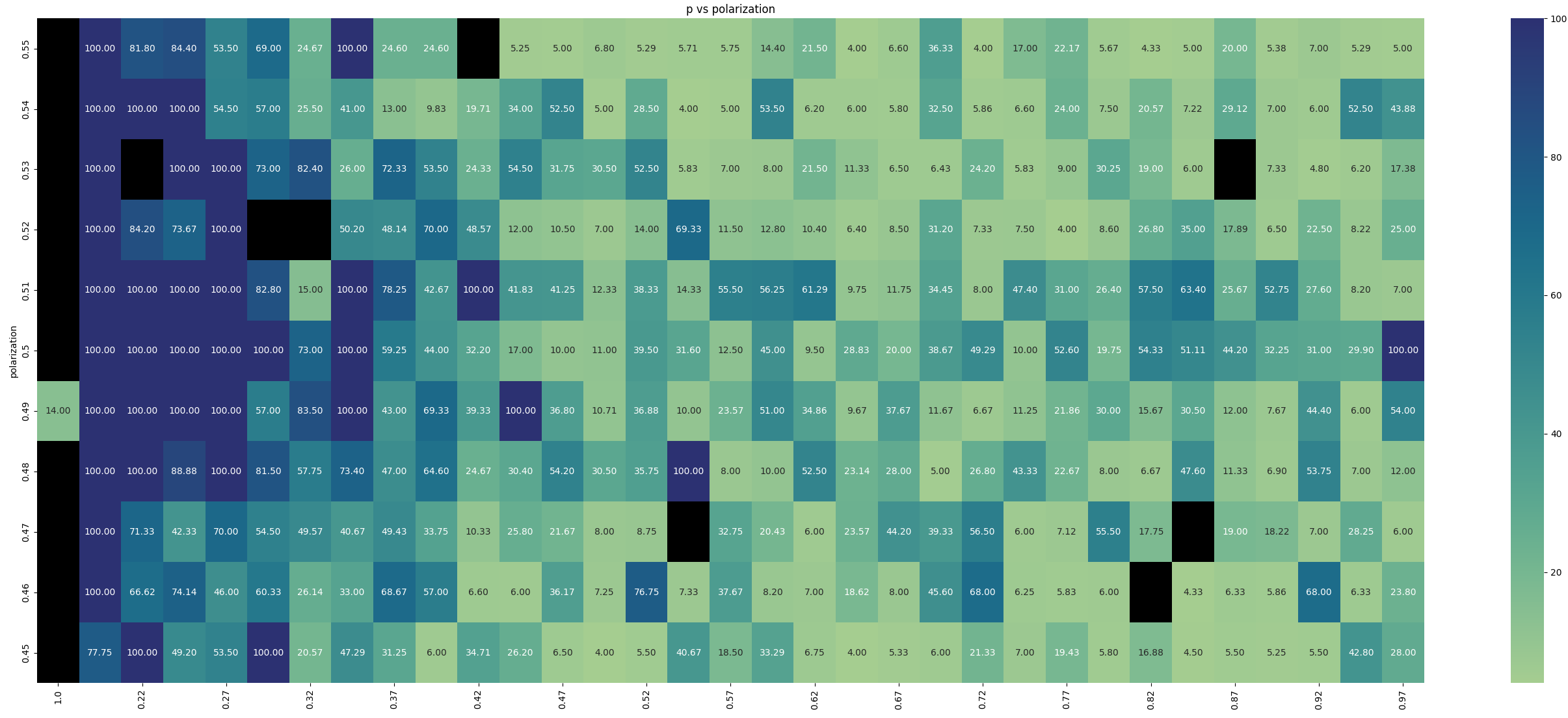}\newline
\quad Here we see a distinct trend towards faster convergence as p increases. Recall that p represented the probability of a connection in a small world graph being randomized; thus, a larger p indicates a greater degree of randomness and, likely, a lower degree of clustering. Therefore, it follows from our observations of Clustering vs Polarization that this map would take the opposite shape. The graphs fail to converge at low p and polarization close to 50\%, and become more likely to converge (and converge more quickly) as we move to the right and away from polarizations close to 50\%.

\subsection{Maximum Degree vs Polarization}
\includegraphics[width=\textwidth]{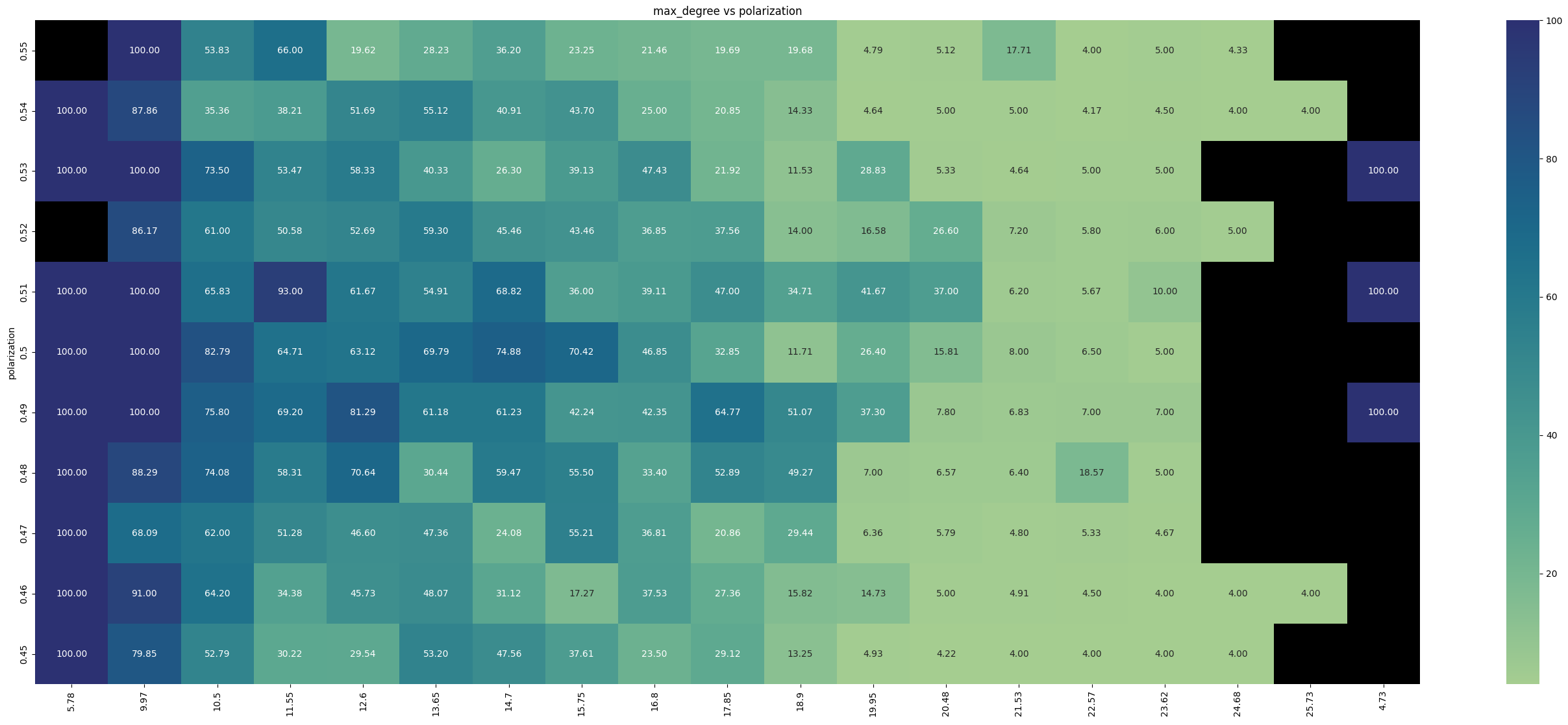}
\includegraphics[width=\textwidth]{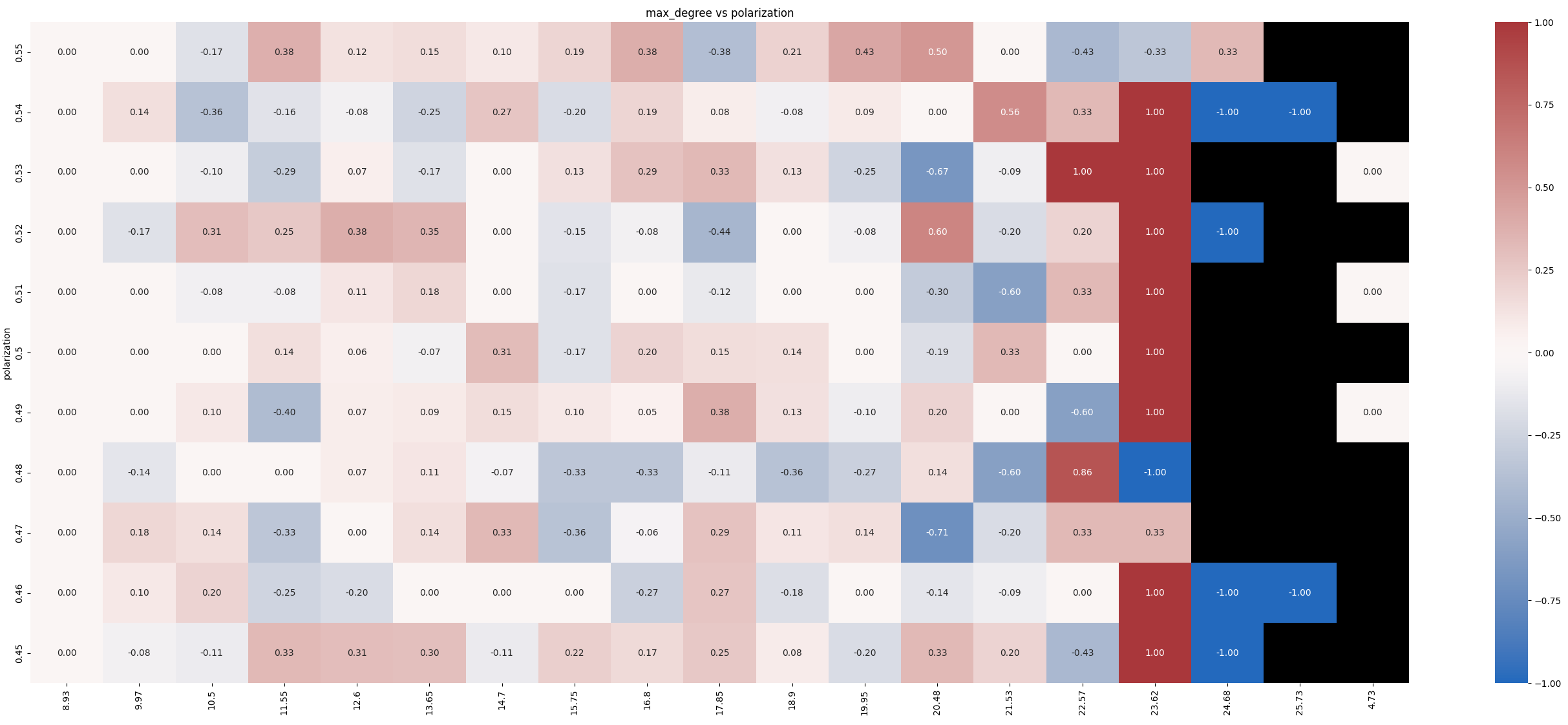}\newline
The first graph indicates that as we increase the maximum degree in the graph – the degree of the node with the most connections --  the rate of converge increases considerably. This suggests that more highly connected graphs converge more quickly, which also aligns with our observations of Density vs Polarization. 
\newline \newline
The second graph indicates more decisive behavior as the maximum degree increases. The tiles are much darker red and blue, meaning that the average color was close to +1 (every graph in the region converged to red) or -1 (every graph in the region converged to blue). This suggests that as a graph becomes more heavily connected, the behavior becomes more consistent or decisive and it becomes more difficult for the graph to be forced to converge to a different color. 
\newline \newline
From the above graphs and observations, we have concluded that graphs converge more quickly at a higher density, more quickly at a lower level of clustering, and have more decisive or consistent behavior at a higher maximum degree. In generating a Watts-Strogatz graph, the k and p can be easily explicitly chosen. Thus, we know how to generate graphs that will converge quickly (or slowly) by manipulating k and p to manipulate density, clustering, and maximum degree. 
\newline \newline
For the following heatmaps, we focused less on the number of states to convergence and more on the color of convergence. 

\subsection{Blue Nodes in Top Ten Most Connected Nodes vs Polarization}
\includegraphics[width=\textwidth]{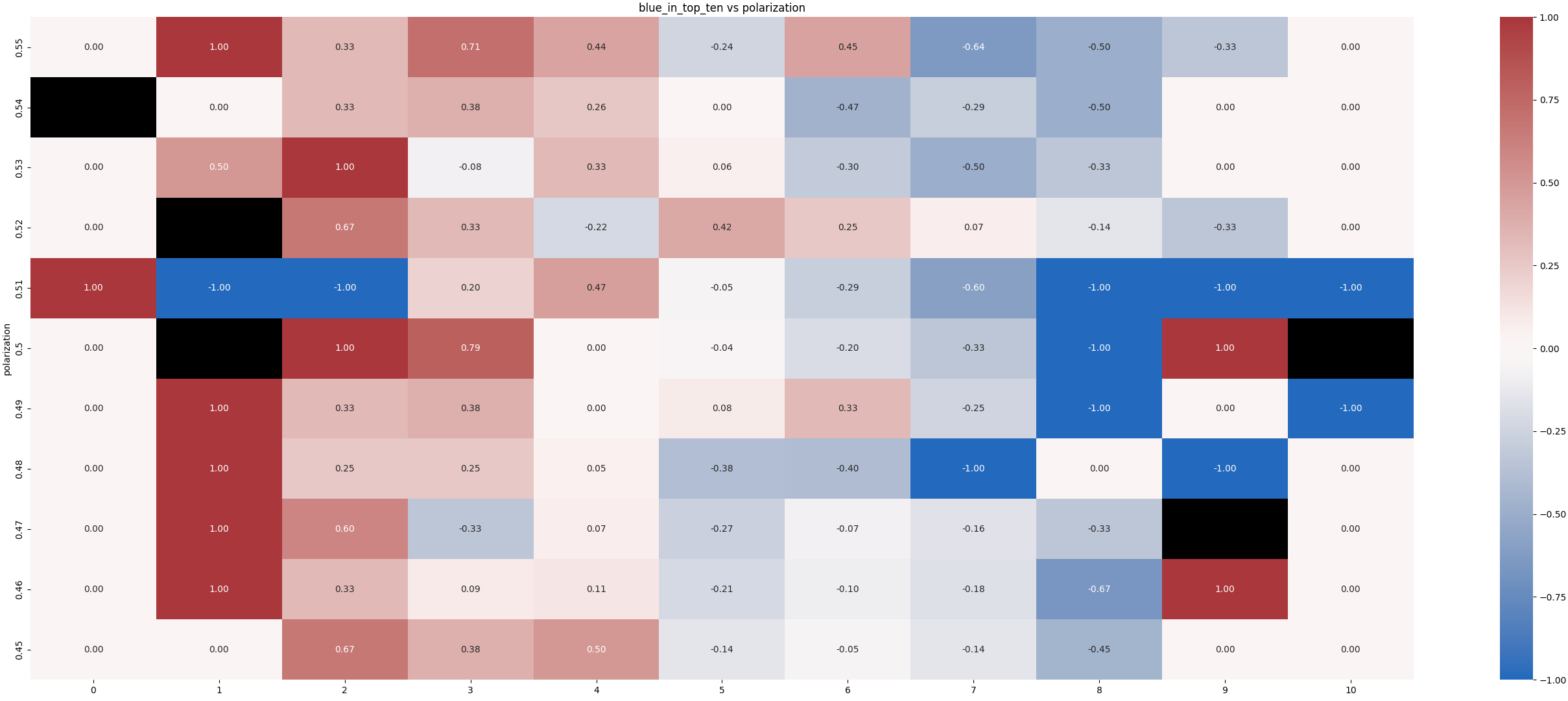}\newline
We can see from this graph that as the number of blue nodes in the top ten most connected nodes increases (and, therefore, the number of red nodes decreases) the graph’s behavior becomes more decisive even close to 50\% polarization. However, the overall trend of the graph seems fairly scattered. Surprisingly, there are more graphs that trend neither blue nor red at the extremes than there are at the center. 

\subsection{Total Degree of Blue Nodes in Top Ten vs Polarization}
\includegraphics[width=\textwidth]{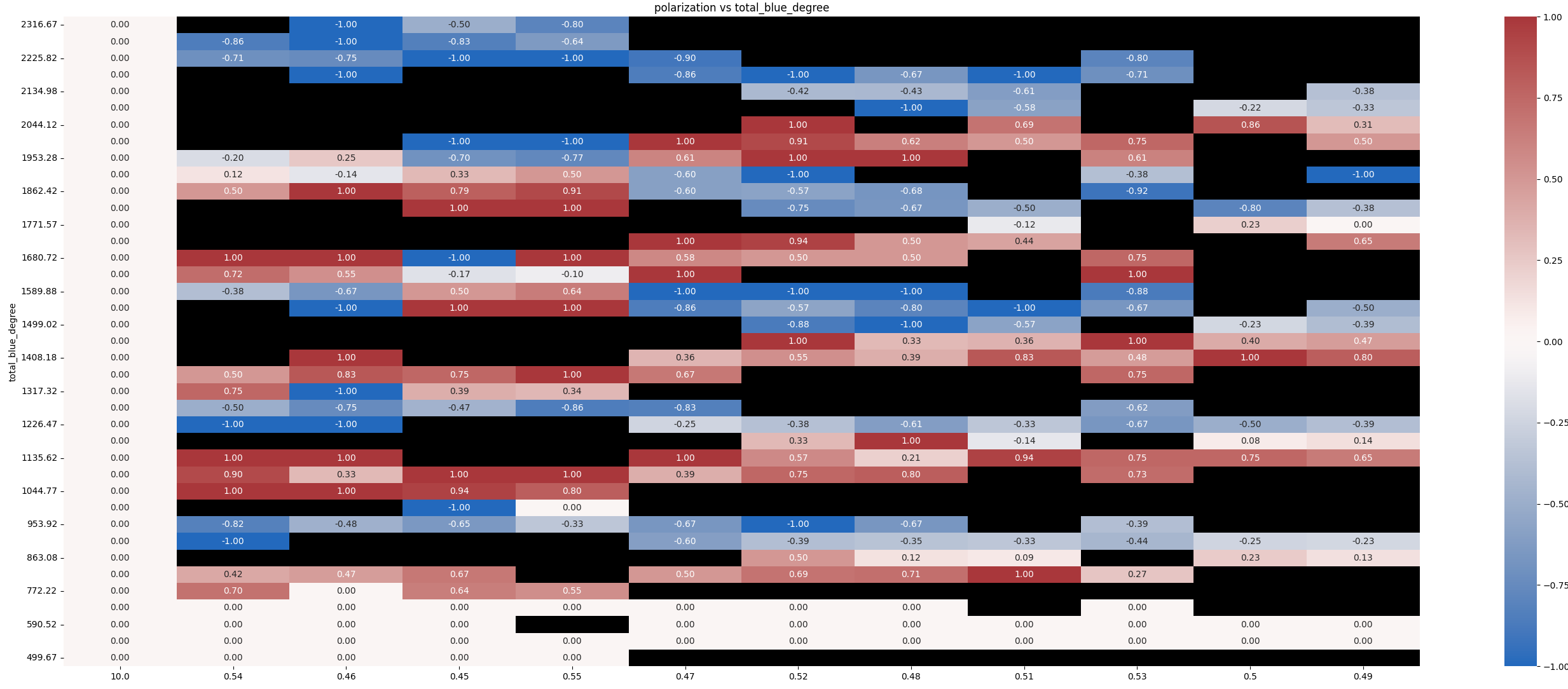}\newline
This graph gives us similar behavior, though there was not much data generated at very high degrees. We can see the mix of blue trending more blue (and more decisively blue) as the total degree of the blue nodes in the top ten most connected nodes increases, and vice versa. Interestingly, this seems to hold true even at polarizations close to 50\%.
\newline \newline 
We also looked at the color of the maximal node, and found that in cases where the node with the highest degree of connection was red the graph converged to red 56.2\% of the time, and in cases where it was blue it converged to blue 56.5\% of the time, indicating that the color of the maximal node is not a very strong predictor of the color of the resultant graph. Additionally, we found that in cases where greater than 50\% of the nodes in the top ten were blue, the graph converged to blue 74.3\% of the time, and in cases where greater than 50\% were red the graph converged to red 74.3\% of the time. This does appear to be a more effective strategy for quickly determining the convergence of a graph. 
\newline \newline
In regards to winnability and the value of the graph, it is likely that graphs that converge quickly are more likely to be winnable, and graphs that fail to converge or converge slowly are not worth investing in. Graphs that behave very decisively are also likely not worth investing in, as they are very likely to converge to whichever color they trend towards regardless of how many resources the player puts forward. Thus, more valauble graphs would be those with high density, low clustering, and a low maximum degree. We also constructed some simple voter strategies, but did not find them to be extremely effective. 

\section{Defining Voter Strategies}
\quad For this project, we identified and studied four voter strategies that can be applied and played against one another. The Random strategy, the Determinist strategy, the Predictive strategy, and the Reactive strategy. 

\subsection{Random Strategy}
This strategy randomly assigns the resources across all battlefields. It was created largely for benchmarking the efficacy of other strategies. 

\subsection{Deterministic Strategy}
For this strategy, we decided to first take into account average degree of connection and polarization, and create a heatmap indicating the number of states to convergence for various combinations of degree and polarization. We also recorded the color that the graph converged to. For each battlefield, this strategy takes a polarization and an average degree of connectedness and looks up the color of convergence and number of states to convergence for the corresponding graph from the heatmap. For each battlefield, its “value” is represented with 1/num\_states, and resources are allocated according to the value of the battlefield. 
\[v = \frac{1}{s}\]
\[r = \frac{v}{(\Sigma(v_i))}*R\]
Where s represents the number of states to convergence, r represents the number of the resources to allocate to a specific battlefield, and R represents the total number of resources available to the player. 
\newline \newline
Put simply. this strategy goes for the "low-hanging fruit." It will win the easiest graphs, but will fail to win any other graphs. Depending on the playing field, it will either beat or tie with the Random Strategy. 

\subsection{Predictive Strategy}
The predictive strategy calculates the behavior of the deterministic strategy and seeks to mold its behavior accordingly. It selects n/2 + 1 battlefields to attempt to win by choosing those battlefields that have the lowest allocation via the determinstistic strategy, and tries to exceed the deterministic strategy’s allocations for these battlefields by a constant amount. The allocations for the n/2 + 1 battlefields are calculated using the following equation. 
\[C = \frac{predictive\, resources - determinstic\, resources}{(n/2)+1}\]
 This strategy can only realistically be played against the Deterministic Strategy, and will win whichever battlefields the Deterministic Strategy ignores, but is unpredictable when both the Determinstic and Predictive Stragies are competing for the same battlefield. 

\subsection{Reactive Strategy}
This is identical to the predictive strategy, but it chooses the battlefields with the highest allocations rather than the lowest allocations. Interestingly, this always or almost always ties with the Deterministic Strategy. 

\section{Conclusion and Further Work}
\quad We have been able to outline a Voter-Blotto Game and isolate some features that impact the behavior of the graph, but we have not yet been able to construct an effective strategy utilizing these features. However, based on the efficacy of existing strategies, it does seem likely that it would be possible to construct a deterministic strategy that takes into account density, clustering, and the degree of the maximal node to assign value to the battlefields and distribute resources accordingly. Some of the features are in conflict with one another – a higher density will likely mean a higher maximal degree, for example, thus they would have to weighed against one another to determine which is more important to the value of the graph. 
\newline \newline 
Additionally, knowing these features, it may also be possible to construct an AI model that decides where to allocate resources based on a Random Forest Regressor (to predict the number of states to convergence) and a Random Forest Classifier (to determine the color of convergence). These will branch based on specified features, so by selecting the relevant features, we may get fairly effective behavior. There are likely other AI models that would apply as well, and including reinforcement learning could help guide us to which features are the most important to value and thus have the biggest impact on the effectiveness of the strategy. 
\newline \newline 
There’s also some additional work that could be done to compare features of the graph. For example, if it were possible to efficiently generate graphs with a constant blue degree and red degree while varying structure, it would be possible to extrapolate more information about graph structures that can outweigh a numerical advantage. However, we currently don’t have a way of efficiently generating graphs with deterministic values for features other than k, p, polarization, and number of nodes. 
\newline \newline
The code and data for this project can be found at https://github.com/rjerome1997/voter-blotto/tree/master

\end{document}